\input harvmac
\noblackbox

\newcount\figno
\figno=0
\def\fig#1#2#3{
\par\begingroup\parindent=0pt\leftskip=1cm\rightskip=1cm\parindent=0pt
\baselineskip=11pt
\global\advance\figno by 1
\midinsert
\epsfxsize=#3
\centerline{\epsfbox{#2}}
\vskip 12pt
{\bf Fig.\ \the\figno: } #1\par
\endinsert\endgroup\par
}
\def\figlabel#1{\xdef#1{\the\figno}}
\def\encadremath#1{\vbox{\hrule\hbox{\vrule\kern8pt\vbox{\kern8pt
\hbox{$\displaystyle #1$}\kern8pt}
\kern8pt\vrule}\hrule}}
\def\apm{{\alpha^{\prime}}}


\font\cmss=cmss10
\font\cmsss=cmss10 at 7pt
\def\rlx{\relax\leavevmode}
\def\inbar{\vrule height1.5ex width.4pt depth0pt}
\def\IC{\relax\,\hbox{$\inbar\kern-.3em{\rm C}$}}
\def\IN{\relax{\rm I\kern-.18em N}}
\def\IP{\relax{\rm I\kern-.18em P}}
\def\ZZ{\rlx\leavevmode\ifmmode\mathchoice{\hbox{\cmss Z\kern-.4em Z}}
 {\hbox{\cmss Z\kern-.4em Z}}{\lower.9pt\hbox{\cmsss Z\kern-.36em Z}}
 {\lower1.2pt\hbox{\cmsss Z\kern-.36em Z}}\else{\cmss Z\kern-.4em
 Z}\fi}
\def\IZ{\relax\ifmmode\mathchoice
{\hbox{\cmss Z\kern-.4em Z}}{\hbox{\cmss Z\kern-.4em Z}}
{\lower.9pt\hbox{\cmsss Z\kern-.4em Z}}
{\lower1.2pt\hbox{\cmsss Z\kern-.4em Z}}\else{\cmss Z\kern-.4em
Z}\fi}
\def\IZ{\relax\ifmmode\mathchoice
{\hbox{\cmss Z\kern-.4em Z}}{\hbox{\cmss Z\kern-.4em Z}}
{\lower.9pt\hbox{\cmsss Z\kern-.4em Z}}
{\lower1.2pt\hbox{\cmsss Z\kern-.4em Z}}\else{\cmss Z\kern-.4em
Z}\fi}

\def\narrowplus{\kern -.04truein + \kern -.03truein}
\def\narrowminus{- \kern -.04truein}
\def\narrowminussub{\kern -.02truein - \kern -.01truein}

\def\half{{1\over 2}}

\def\m{{\mu}}
\def\n{{\nu}}

\def\d{{\delta}}
\def\t{{\theta}}
\def\a{{\alpha}}
\def\T{{\Theta}}
\def\frac#1#2{{#1\over #2}}

\def\o{{\rm ord}}

\def\IZ{\relax\ifmmode\mathchoice
{\hbox{\cmss Z\kern-.4em Z}}{\hbox{\cmss Z\kern-.4em Z}}
{\lower.9pt\hbox{\cmsss Z\kern-.4em Z}}
{\lower1.2pt\hbox{\cmsss Z\kern-.4em Z}}\else{\cmss Z\kern-.4em
Z}\fi}
\def\IB{\relax{\rm I\kern-.18em B}}
\def\IC{{\relax\hbox{$\inbar\kern-.3em{\rm C}$}}}
\def\ID{\relax{\rm I\kern-.18em D}}
\def\IE{\relax{\rm I\kern-.18em E}}
\def\IF{\relax{\rm I\kern-.18em F}}
\def\IG{\relax\hbox{$\inbar\kern-.3em{\rm G}$}}
\def\IGa{\relax\hbox{${\rm I}\kern-.18em\Gamma$}}
\def\IH{\relax{\rm I\kern-.18em H}}
\def\II{\relax{\rm I\kern-.18em I}}
\def\IK{\relax{\rm I\kern-.18em K}}
\def\IP{\relax{\rm I\kern-.18em P}}
\def\p{\partial}
\font\cmss=cmss10 \font\cmsss=cmss10 at 7pt
\def\IR{\relax{\rm I\kern-.18em R}}

%

%
%
\def\eqnn#1{\xdef #1{(\secsym\the\meqno)}\writedef{#1\leftbracket#1}%
\global\advance\meqno by1\wrlabeL#1}
\def\eqna#1{\xdef #1##1{\hbox{$(\secsym\the\meqno##1)$}}
\writedef{#1\numbersign1\leftbracket#1{\numbersign1}}%
\global\advance\meqno by1\wrlabeL{#1$\{\}$}}
\def\eqn#1#2{\xdef #1{(\secsym\the\meqno)}\writedef{#1\leftbracket#1}%
\global\advance\meqno by1$$#2\eqno#1\eqlabeL#1$$}

\lref\sw{N.
Seiberg and E. Witten, ``String Theory and Noncommutative
Geometry'', hep-th/9912072.} \lref\berg{E.~Bergshoeff,
D.~S.~Berman, J.~P.~van der Schaar and P.~Sundell, ``A
noncommutative M-theory five-brane,'' hep-th/0005026.}

\lref\hklm{
J.~A.~Harvey, P.~Kraus, F.~Larsen and E.~J.~Martinec,
``D-branes and strings as non-commutative solitons,''
hep-th/0005031.}
\lref\gms{R.~Gopakumar, S.~Minwalla and A.~Strominger,
``Noncommutative solitons,''
JHEP {\bf 0005}, 020 (2000)
[hep-th/0003160].} 
\lref\wsft{E.~Witten,
``Noncommutative Geometry And String Field Theory,''
Nucl.\ Phys.\  {\bf B268}, 253 (1986).} 
\lref\romans{L.~J.~Romans,
``Operator Approach To Purely Cubic String Field Theory,''
Nucl.\ Phys.\  {\bf B298} (1988) 369.} 
\lref\tachvac{ A.~Sen and  B. Zwiebach, 
``Tachyon condensation in string field theory,''
JHEP { \bf 0003} (2000) 002.} 
\lref\moe {N.~Moeller and  W.~Taylor, ``
Level truncation and the tachyon in open bosonic string field theory,''
hep-th/0002237 .}
\lref\senb{ A.~Sen, ``Descent Relations Among Bosonic D-branes,''
 Int.J.Mod.Phys. { \bf A14} (1999) 4061-4078, hep-th/9902105.}
\lref\evid{J.~A.~Harvey and P.~Kraus,
``D-branes as unstable lumps in bosonic open string field theory,''
JHEP {\bf 0004}, 012 (2000)
[hep-th/0002117].} 
\lref\jev{R.~de Mello Koch, A.~Jevicki, M.~Mihailescu and R.~Tatar,
``Lumps and p-branes in open string field theory,''
Phys.\ Lett.\  {\bf B482}, 249 (2000)
[hep-th/0003031].}
\lref\evidb{N.~Moeller, A.~Sen and B.~Zwiebach,
``D-branes as tachyon lumps in string field theory,''
hep-th/0005036.}
\lref\rms{K.~Dasgupta, S.~Mukhi and G.~Rajesh,
``Noncommutative tachyons,''
JHEP {\bf 0006}, 022 (2000)
[hep-th/0005006].}
\lref\senc{A.~Sen,
 ``Universality of the tachyon potential,''
JHEP {\bf 9912} (1999) 027
[hep-th/9911116].}
\lref\send{A.~Sen,
``Supersymmetric world-volume action for non-BPS D-branes,''
JHEP {\bf 9910}, 008 (1999)
[hep-th/9909062].}
\lref\gar{M. Garousi, ``Tachyon couplings on non-BPS D-branes and 
Dirac-Born-Infeld action'', hep-th/0003122.}
\lref\berg{E.~A.~Bergshoeff, M.~de Roo, T.~C.~de Wit, E.~Eyras and S.~Panda,
``T-duality and actions for non-BPS D-branes,''
JHEP {\bf 0005}, 009 (2000)
[hep-th/0003221].}
\lref\klu{J.~Kluson,
``Proposal for non-BPS D-brane action,''
hep-th/0004106.}
\lref\cubic{
G.~T.~Horowitz, J.~Lykken, R.~Rohm and A.~Strominger,
``A Purely Cubic Action For String Field Theory,''
Phys.\ Rev.\ Lett.\  {\bf 57}, 283 (1986).}
\lref\shenker{S.~Shenker, ``What are Strings Made of?'' 
talk given at ``String Theory at the Millenium'',
http://quark.theory.caltech.edu/people/rahmfeld/Shenker/fs1.html}
\lref\sig{See for example, W. Seigel, 
``Covariantly Second Quantized String.2; 3,'' Phys.\ Lett.\  {\bf B149}, 157, 162 
(1984); {\bf B151}, 391, 396, (1984); T. Banks and M. Peskin, 
``Gauge Invariance of String Fields'' 
Nucl. Phys. {\bf B264}, 513 (1986). } 
\lref\rast{L.~Rastelli and B.~Zwiebach,
``Tachyon potentials, star products and universality,''
hep-th/0006240.}
\lref\schom{V.~Schomerus,
``D-branes and deformation quantization,''
JHEP {\bf 9906}, 030 (1999)
[hep-th/9903205].}
\lref\kost{V.~A.~Kostelecky and S.~Samuel,
``The Static Tachyon Potential In The Open Bosonic String Theory,''
Phys.\ Lett.\  {\bf B207}, 169 (1988).}
\lref\padic{D.~Ghoshal and A.~Sen,
``Tachyon condensation and brane descent relations in p-adic string  theory,''
hep-th/0003278.}
\lref\witnew{E. Witten, ``Noncommutative Tachyons and String Field 
Theory'', hep-th/006071.}

\Title
{\vbox{
\baselineskip12pt
\hbox{hep-th/0007226}\hbox{}\hbox{}
}}
{\vbox{
\centerline{Symmetry Restoration and Tachyon Condensation}
\centerline{in Open
String Theory}
}}

\centerline{ Rajesh ${\rm Gopakumar}$, Shiraz ${\rm Minwalla}$,
and Andrew ${\rm Strominger}$}
\bigskip\centerline{ Jefferson Physical Laboratory}
\centerline{Harvard University} 
\centerline{Cambridge, MA 02138}
\smallskip

\vskip .3in \centerline{\bf Abstract} {It has recently been argued that 
D-branes in bosonic string theory 
can be described as noncommutative solitons, 
outside whose core the tachyon is condensed to its 
ground state. We conjecture that, in addition, the local $U(1)$ 
gauge symmetry is restored to a $U(\infty)$ 
symmetry in the vacuum 
outside this core. We present new solutions obeying this 
boundary condition. The tension of these solitons 
agrees exactly  with the 
expected D-brane tension for arbitrary 
noncommutativity parameter 
$\t$, which effectively becomes a dynamical variable. 
The restored $U(\infty)$ eliminates unwanted extra modes 
which might otherwise appear outside the soliton core. 
}
\smallskip
\Date{}
\listtoc
\writetoc
\newsec{Introduction}
 The general theory of relativity follows largely from the demand
 that the laws of physics take the same form in all coordinate
 systems. In string theory, the massless boson associated to this
 coordinate invariance - namely the graviton - is just one mode of
 an infinite tower of mostly massive string states. Associated to this
 infinite tower of modes is a stringy generalization
 of coordinate invariance. In the usual perturbative string
 vacuum, almost all of the string modes are massive and almost
 all of this stringy symmetry is accordingly spontaneously broken \sig .

One may expect that string theory itself largely follows from the
demand of stringy symmetry. However, despite the spectacular
developments of the last five years, the nature of this stringy
symmetry remains enigmatic.\foot{A recent discussion can be found
in \shenker.} In this paper we investigate 
this issue of (open) stringy symmetry restoration in 
the context of a recent circle
of ideas involving tachyon condensation, D-branes and
noncommutative geometry. We will consider only the 
classical\foot{Quantum effects could well be important in tachyon
condensation, but we will not consider them.} open bosonic string.
 
Following the work of Sen \refs{\senb, \senc}, 
it is widely believed that the endpoint of the condensation of the 
open string tachyon is the closed string vacuum.
There is by now compelling evidence for this conjecture from diverse 
points of view, including numerical computations \refs{\tachvac, \moe, \rast}
using Witten's open string field theory \wsft. 
Moreover, Sen has argued \senb\ 
(see also \refs{\evid, \jev, \evidb})
that  D-branes in bosonic string theory can
be viewed as solitons of the open string tachyon. Outside the core
of the soliton the tachyon is in its ground state, and the theory is 
in the closed string vacuum with no open string excitations. 

Recently Harvey, Kraus, Larsen and Martinec \hklm\ and 
Dasgupta, Mukhi and Rajesh \rms\ have shown that 
turning on a large $B$ field enables an elegant realization 
of D-branes as tachyonic solitons. Techniques from
noncommutative field theory  \gms\ can be used to construct the
D-brane soliton in the $\t \to \infty$ limit of large noncommutativity.
The soliton and D-brane tensions agree exactly in this limit. A
simple and beautiful explanation of the non-abelian
structure of D-branes is found 
\hklm, with a natural embedding into string field
theory \witnew.

However, even with these improvements 
several puzzles remain. 
In order to eliminate
unwanted propagating open string states 
far outside the D-brane soliton core (i.e. in the closed string vacuum), 
one must assume that the coefficients in the tachyon-Born-Infeld Lagrangian 
take special values together with a special choice of field variables. 
Even with these assumptions, unwanted propagating
modes persist inside the core
in the bifundamental of $U(N)\times U(\infty -N)$, where $N$ is the number
of D-branes.
Although plausible mechanisms \refs{ \kost, \padic} 
for the elimination of these modes 
have been proposed, it is unsatisfying that these  
depend on unknown higher stringy corrections and cannot be seen directly
from 
the Lagrangian employed in the analysis. 
In addition it is difficult to understand why ${1
\over \t}$ corrections would not spoil the exact agreement found in \hklm\
between the soliton and D-brane tensions.

In this paper we consider the open bosonic string theory 
in the presence of a maximal rank $B$ field. 
We propose that in the process of tachyon condensation, 
as the tachyon rolls to its minimum, the noncommutative gauge field  
simultaneously rolls to a maximally symmetric configuration, about
which the noncommutative gauge symmetry is fully 
unbroken, and becomes a linearly realized $U(\infty )$.\foot{ We will refer to a configuration as having
unbroken gauge symmetry if all fields  
are left invariant by the gauge
transformations. Note that, with this usage, the usual perturbative
vacuum of a gauge theory breaks local gauge invariance as $\d A \neq 0$
for non-constant gauge transformations.} 
The propagation of open string modes in this `nothing' state
is forbidden by the $U(\infty)$ symmetry, and there is
no need to invoke higher-order stringy corrections or special values 
of coefficients for their elimination.

We also modify the proposed identification of D-branes as
noncommutative tachyon solitons by demanding that far from the core of the
soliton, the solution approaches the nothing state, in which the 
$U(\infty)$ symmetry of the noncommutative field
theory (which is broken to a local $U(1)$ on the D-brane) is
completely unbroken.
We construct exact soliton solutions of the 
noncommutative tachyon-Born-Infeld Lagrangian 
obeying the modified boundary conditions, 
without expanding in ${1 \over \theta}$. 
It is further argued that these are exact-to-all-orders solutions 
of classical open string theory. The soliton tension
exactly matches the expected D-brane tension. Furthermore, the propagation
of open string modes far from the core, (i.e. in the nothing state)
is forbidden as above by the $U(\infty)$ symmetry.
We regard these successes as evidence for the conjecture that the 
vacuum outside
the D-brane core is the state of fully unbroken open string symmetries.
On the other hand the situation for the bifundamentals is somewhat
improved,
but not fully resolved, as will be discussed in section 4.1. 

One way of understanding the $\t$-independence of the D-brane tension is 
that, in the context of tachyon condensation, $\t$ is effectively a dynamical
variable. In a sense (to be made precise herein), our proposal is that $\t$ 
effectively relaxes to $\infty$  at the boundary. 

While some puzzles are resolved in our approach, a significant new puzzle
arises. In addition to the solutions corresponding to 
D-branes, there are a number of other spurious 
solutions obeying the same boundary
conditions for which we have no
physical interpretation. These must be understood or somehow excluded 
before the picture presented here can be
regarded as complete.

\newsec{The Action in Shifted Variables}

\subsec{The Action}

The Euclidean action for
$U(1)$ open bosonic string theory contains the terms \refs{\send,\senc}
(see also \refs{\gar,\berg,\klu})
\eqn\fti{S={1 \over G_o^2\apm^{13}(2\pi)^{25}}\int d^{26}x
\left( V(T)\sqrt{{\rm det}(G+2\pi\apm F)}+{\apm \over 2}f(T)D_\m TD^\m
T\sqrt{{\rm det}G}+\cdots \right).}
The tachyon potential $V$ has a maximum at $T=T_{max}$
corresponding to the unstable perturbative string vacuum and a
minimum at $T=T_{min}$ which should contain no perturbative open
string excitations. According to \senb\ the  minimum obeys
\eqn\mbc{V(T_{min})=0,} and the maximum is determined from the D25-brane
tension to be, in our conventions \eqn\pfs{V(T_{max})= 1.}
The universal coefficient of the potential term in \fti\ was 
demonstrated with worldsheet methods in \senb. 
In addition it has been conjectured that $f(T_{min})=0$ \refs{\gar,\berg}.
\foot{Note that 
if $f$ is smooth at $T_{min}$ it can in any case be set to
one by a field redefinition.} This will not 
play an essential role in our analysis, although it is 
required in \hklm. 

We wish to study the open bosonic string theory in the background of 
a constant $B$ field. According to \refs{\schom, \sw}, 
the Euclidean action in this background 
continues to be given by \fti, except that: 
\item{a.} Space becomes noncommutative, i.e. all  
products in \fti\ are replaced by star products, with a noncommutativity
tensor $\T$, whose value is given below. 
\item{b.} The parameters that appear in \fti; the open string 
metric $G_{\m\n}$, the noncommutativity tensor $\T^{\m\n}$
and the open string coupling $G_o$ are related to closed string 
moduli by the formulae
\eqn\gya{\eqalign{2\pi \apm & {G}^{\m \n }+\T^{\m \n}
=\bigl({ 2\pi\apm\over g+2\pi
\apm B}\bigr)^{\m \n},\cr
              G_o^2&=g_{str}\sqrt{{\rm det}(g+2\pi
\apm B) \over {\rm det}  g}.}} 
Here $g$ and $g_{str}$ are the usual constant closed string metric 
and coupling.

We are thus led to study a $U(1)$ noncommutative gauge theory, 
interacting with a scalar field (the tachyon) that transforms in the 
adjoint of the gauge group. Note that we have not taken the 
$\apm \to 0$ scaling limit,  so Born-Infeld corrections
are retained. The noncommutative action \fti\ together with \gya\
is identical to that considered in \hklm\ 
(prior to taking the $\t \to \infty$ limit). 
An alternate form of the action, used for example in 
\refs{\rms,\gar,\berg,\klu} differs by higher derivative tachyon 
terms which would not affect our conclusions. 

We choose $B_{\m\n}$ so that space is maximally noncommuting, i.e.
$\T$ has maximal rank. We parameterize space  with complex
coordinates $z^m$, $m=1,...13$ obeying 
\eqn\tyh{[z^m, \bar z^{\bar n}]= i\T ^{m\bar n}.}

\subsec{Brief Review of the Operator Formalism}

In this subsection we recall certain facts about noncommutative 
field theories, and especially noncommutative gauge
theories, that we will use in our construction below. 
See, for instance, \gms\ for more details. 
   
The algebra of functions on a 26 dimensional noncommutative space 
is represented by operators on the Hilbert space of a thirteen 
dimensional particle.  On this space, we define thirteen 
annihilation operators $a_{{\bar m}}$ and an equal number of 
creation operators  $a^{\dagger}_m$
\eqn\creann{a_{{\bar m}}=-i\T_{{\bar m} n}^{-1} {z}^{n}, ~~~
a^{\dagger}_m=i\T_{m{\bar n}}^{-1} {\bar z}^{\bar n}.}
These operators obey the commutation relations
\eqn\xtyh{[a^\dagger_m, a_{\bar n}]= -i\T ^{-1}_{m\bar n}.} 
Several useful relations in translating from functions to operators are
\eqn\usf{\eqalign{\int d^{2n}x &\to (2\pi)^n\sqrt{ (-)^n{\rm det} \T}{\rm
Tr},\cr
\p_m &\to -[a_m^\dagger,~~],\cr 
\p_{\bar m} &\to [a_{\bar m},~~].}}
We now consider a noncommutative gauge theory written in the operator 
language. The covariant derivative of a field $\ph$ that transforms in the 
adjoint of the noncommutative gauge group may be cast in the form
\eqn\coder{ D_m \ph = \p_m \ph +i[A_m, \ph]=-[C_m, \ph]; ~~~~
D_{\bar m} \ph = \p_{\bar m}\ph+i[A_{\bar m}, \ph]=[C_{\bar m}, \ph]}
where 
\eqn\fsss{C_m=-iA_m+{a^\dagger_m
},~~~~~~~C_{\bar m}=iA_{\bar m}+{a_{\bar m} }.}
The noncommutative field strength is
\eqn\ncfs{{F}_{m \bar n}=i[C,\bar C]_{m \bar n}-\T^{-1}_{m \bar n}}
where $[C,\bar C]_{m \bar n}= [C_m,C_{\bar n}]$ and 
$\T^{-1}_{m \bar n}\T^{ \bar n p}=\delta_m^{~~p}.$
The fields $C_m$, $C_{\bar m}$ transform homogeneously under gauge 
transformations. In particular, the field configurations 
$C_m=C_{\bar m}=0$ leave the gauge symmetry unbroken. 

\subsec{The Action and Equations of Motion}

The noncommutative action 
\fti\ (for open string modes in the presence of a $B_{\m\n}$ field)
can be rewritten in operator language as 
\eqn\rfgh{\eqalign{S=&{\sqrt{-{\rm det} \T} \over G_o^2
\apm^{13}(2\pi)^{12}}
{\rm
Tr}\Bigg[V(T)\sqrt{{\rm
det} (G+2\pi\apm (i[C,\bar C]-\T^{-1}))}
\cr
&+\apm f(T)[C_p,T][T,C^p]\sqrt{{\rm det}(G)} + \cdots \Bigg].}}
Operators in \rfgh\ are appropriately 
ordered so as to reproduce string amplitudes; 
the precise ordering of operators in this action 
will not be important for us.

The tachyon equation of motion that follows from \rfgh\ is 
\eqn\teom{
2\apm f(T)[C_m,[C^m,T]]\sqrt{{\rm det}G}-\apm f'(T)[C_m,T][C^m,T]
+V'(T)\sqrt{{\rm det}M}=0,}
where we have defined 
\eqn\mdf{M=G+2\pi\apm (i[C,\bar C]-\T^{-1}).}  The
equation for $C_m$ is 
\eqn\gti{ -\half \apm [T,[C^m,T]f(T) ]\sqrt{{\rm
det}G}+i\pi \apm [C_{\bar n},V(T)
\sqrt{{\rm det}M}(M^{-1})^{m \bar n}] =0.}

\newsec{The Nothing State}
 
In the variables \fsss, the usual vacuum with a single
D25-brane is \eqn\fgg{T=T_{max},~~~~~C_m= {a^\dagger_m }.} In \senc\ it was
conjectured 
that the `nothing' state with no D25-branes is
\eqn\dfgg{T=T_{min},~~~~~C_m= {a^\dagger_m }.} We would like to
propose instead that the nothing state is
\eqn\fgag{T=T_{min},~~~~~C_m=0.} Due to \mbc , \fgag\
and \dfgg\ are energetically degenerate.

Under a local $U(1)$ gauge transformation
\eqn\rtj{\delta A_m =\p_m\epsilon + i [A_m,\epsilon],}
it follows from 2.7 that $C_m$ transforms as 
\eqn\ratj{\delta C_m =i [C_m,\epsilon].}
Hence, as remarked above,  the nothing state \fgag\ is fully invariant 
under this symmetry. The local $U(1)$ symmetry 
is restored to an unbroken $U(\infty)$ 
symmetry of unitary transformations on the quantum mechanical Hilbert 
space. 

We will now argue that fluctuations about 
the fully symmetric state \fgag\ have no perturbative 
propagating open string degrees of freedom. \fti\ describes a 
noncommutative gauge theory whose matter fields all 
transform in the adjoint of the gauge group. Gauge invariance 
dictates that derivatives and gauge fields $A$ appear in the action  
only in the combination $C_m$. Thus fluctuations about 
any background with $C_m=0$ are governed by an action with no 
derivatives beyond those that appear in the star product. 
In particular, quadratic terms in the action have no derivatives
(as the star product acts trivially on such terms); this 
statement is unchanged by nonlinear field redefinitions.  
Further, $U(\infty)$ invariance ensures that 
explicit derivative terms are not dynamically generated.
Thus, perturbatively, open string modes do not propagate about the
background \fgag.

\subsec{Nothing in Ordinary Variables}

In \sw\ it was shown that there is a nonlocal 
field redefinition which relates the noncommutative 
field strength $F$ to an `ordinary'  field strength,
which we shall denote $F^\o$, appearing in the commutative 
formulation of the same theory. Under this Seiberg-Witten map
\sw, a constant noncommutative field ${F}$ maps to a constant 
ordinary field strength, whose value is given by 
\eqn\swmap{ F^\o= {F}{1 \over 1 + \Theta {F}}.}
In \fgag\ the noncommutative 
field strength $F$  takes the constant
value $-\Theta^{-1}$, and so corresponds to a divergent ordinary
field strength\foot{We
are grateful to Jeff Harvey for explaining this point.}. Schematically, 
\fgag\ in ordinary variables, and the gauge $B=0$ takes the form
\eqn\ordconf{T=T_{min}~~~~~F^\o_{\m\n}=\infty.}
By the second equation in \ordconf\ we mean that, 
in the vacuum state,  $F^\o$ is a rank 26 tensor, all of whose 
eigenvalues diverge.

Thus the conjecture of this section may be worded in ordinary variables
as follows: The perturbative 
open string vacuum 
state with $T=T_{max}$, and a finite constant $F^\o$  (in the 
gauge $B=0$) is unstable
and decays to the nothing state, $T=T_{min}$ and infinite $F^\o$.
This is in part possible because at $T=T_{min}$,  $V(T)$ vanishes and hence 
there is no energy cost to changing $F^\o$.

\subsec{Nothing as $\T=\infty$}

We have argued above that in  
ordinary variables the nothing state is given by  \ordconf, 
or equivalently by
\eqn\ordconftr{T=T_{min},~~~~~F^\o_{\m\n}=0,~~~~~B_{\m\n} \to \infty.}
In order to analyze this state, we move to yet another set 
of variables; the gauge field whose noncommutativity
is set by the large $B$ of \ordconftr.
In terms of the new noncommutative $F$
(whose background value is zero in the nothing state), the action 
takes the form \fti\ with parameters
\eqn\param{G_{\m\n}=-(b {1 \over g} b)_{\m\n},~~~ 
\T^{\m\n}={ 2\pi \apm} ({1\over b})^{\m\n},~~~ G_o^2=g_{str}
\sqrt{ {\rm det} b \over {\rm det} g },}  
where $b_{\m\n}=2 \pi \apm B_{\m\n}$.   
Notice that $\T^2 \equiv \Tr( \T G \T G)=(2\pi \apm)^2\Tr 
( {1 \over g} b {1 \over g} b) \to \infty$
in the limit under consideration. Thus, focusing on energies
for which noncommutative phases are finite, 
explicit derivatives in the action \fti\ may be dropped. 
Defining a rescaled gauge field 
$H_{\m}= g_{\m \a}({1 \over b})^{\a \n}A_{\n}$
\fti\ takes the form 
\eqn\ftip{\eqalign{S={1 \over G_o^2 \apm^{13}(2\pi)^{25}}\int d^{26}x
&\sqrt{ \det G  }
\Bigg( V(T) \sqrt{{\rm det}(\d^\n_\m+2\pi i \apm[H_{\m}, H_{\a}] g^{\a\n})} \cr
+&{ \apm f(T) \over 2} [H_\m, T][T, H_\n]g^{\m\n}+ \cdots )\Bigg). \cr}}
In the operator language 
\eqn\ftipo{S={2 \pi \over g_{str}}\Tr
\left( V(T) \sqrt{{\rm det}(\d_{\m}^{\n}+2\pi i \apm[H_{\m}, H_{\a}] g^{\a \n})}
+{ \apm f(T) \over 2} [H_\m, T][T, H_\n]g^{\m\n}+ \cdots \right), }
where we have used 
\eqn\algident{{ \sqrt{ {\rm det \T}} \sqrt{{\rm \det } G} \over (2 \pi)^{13} 
\apm^{13} G_o^2}={1 \over g_{str}}.}

Thus fluctuations about the nothing state are governed by 
the action \ftipo, 
the dimensional reduction of the infinite $N$ 
open string field theory to a spacetime point.
Note that $B$ does not enter into \ftipo, consistent with the  
expectation that the end product of tachyon condensation is insensitive to the 
initial value of $\T$.

\newsec{D23 Branes}
\subsec{The Soliton Solution}
We wish to find a soliton solution which is translationally
invariant in 24 directions and approaches the nothing state in
the complex transverse $z^{1}$ direction  away from the core. For
these purposes we take 
\eqn\ffz{\eqalign{G^{1 \bar i}=\T^{1 \bar i}&=0,~~~i=2,...13,\cr
\T^{1\bar 1}&= \t G^{1 \bar 1} .}}

Consider the field configuration \eqn\soln{\eqalign
{T-T_{min}&=(T_{max}-T_{min})P_{N_1},\cr C_i&=P_{N_2}a^\dagger_i,~~~~~~~
i=2,...13,\cr C_{1}&=0, }} where $P_{N_k}$ is a 
rank $N_k$ projection
operator in the Hilbert space constructed from $a^\dagger_{1}$.
For example we could take $P_{N_k}$ to be the projection onto the first 
$N_k$ states of the harmonic oscillator. Then the right hand side 
of \soln\ vanishes exponentially outside the soliton core, and the 
solution is asymptotic to the nothing state \fgag.\foot{We note that 
in the $\t \to \infty$ limit this solution is of the general form required for 
the string field theory construction described in \witnew.} 
(In contrast, the approximate solutions found in \refs{\rms,\hklm} 
have the same tachyon field but $C_m=a_m^\dagger$, and are asymptotic to 
\dfgg. )

It is easy to check that \soln\ solves the equations of motion
\teom, \gti. 
The first term in the tachyon equation \teom\ vanishes if we require
\eqn\oft{[P_{N_1}, P_{N_2}]=0.} The second term vanishes because
$V'(T_{max})=0$. \oft\ also implies the separate vanishing of both terms 
in the $C^m$ equation \gti. 

Of course, the true equations of motion that follow from the action \rfgh\
have an infinite number of terms (from the $\cdots$ in \rfgh) that 
we have not considered here. However, each of these terms contains 
at least one factor of a covariant derivative of either $T$ or ${F}$. 
Since all covariant derivatives of $T$ and $F$ given in \soln\
vanish, additional terms in the equation of motion also vanish
to all orders in the $\apm$ expansion. Non-perturbative effects
could alter the situation. It is rather surprising that 
an exact-to-all-orders solution can be constructed without 
even knowing what the Lagrangian is! Usually such constructions are 
possible only with supersymmetry: here it is a consequence of the magic of 
noncommutativity. 

\subsec{The Soliton Action}
We will interpret solutions of the form \soln\ with  
\eqn\rrtg{ P_{N_1}=P_{N_2}=P_N}
as $N$ coincident D-branes. Solutions of the Lagrangian \fti\ 
with $P_{N_1}\neq P_{N_2}$ certainly exist but they 
do not correspond to conventional D-branes (the spectrum is
wrong).  The role of these 
solutions - or a rationale for their exclusion - must
be understood before the picture presented here can be regarded as
satisfactory. For now we consider \rrtg.
Using \ffz, \rfgh\ reduces to  
\eqn\rffc{\eqalign{S&={ V(T_{max}) {\rm Tr}P_N  \over G_o^2 
\apm^{13}(2\pi)^{12}}\sqrt{-{\rm det}\T}
\sqrt{{\rm det}G}\sqrt{1+\bigl({2\pi\apm \over \t }\bigr)^2}.
\cr } }
We wish to rewrite this in terms of 
the coupling ($G_o'$), measure ($\sqrt{{\rm det}G'}$) and noncommutativity 
parameter ($\T^{'i\bar j}$) with respect to the 
24 longitudinal dimensions. It follows from \gya\ that 
these are related to the 26-dimensional 
quantities by
\eqn\pkxc{\eqalign{G_o^2&= {{G_o'}^2  
\sqrt{1 + \bigl({\t \over 2 \pi \apm} \bigr)^2}},\cr \sqrt{ {\rm det}G}&=
G_{1\bar 1}\sqrt{ {\rm
det}G'},
\cr  \sqrt{-{\rm det}\T}&= \t G^{1\bar 1}\sqrt{{\rm det}\T'}.}}
The trace gives
\eqn\fdx{ {\rm Tr}P_N ={N V_{24}\over \sqrt{ {\rm det}G'}(2\pi)^{12} \sqrt{{\rm
det}\T'}},}
where $V_{24}=\int d^{24} y\sqrt{ {\rm det}G'}$. 
Substituting into \rffc\ and using \pfs\ yields 
 \eqn\sdfr{ S= {N V_{24}\over G_o^{'2} \apm^{12}(2\pi)^{23}}.}
All $\t$ dependence has disappeared, and this is 
exactly the action of $N$ parallel D23-branes.

\subsec{The Spectrum}
We now describe the spectrum of the solution \soln\ \rrtg. 
We choose a basis in which
\eqn\bvx{P_N=\sum_{a=1}^N |a \rangle \langle a|.}
\vskip.2in
\noindent{$U(N)$ Adjoint Fields}

$U(N)$ adjoint fluctuations in the tachyon field can be expanded as 
\eqn\trex{\delta T=\sum_{a,b=1}^N T_{ab}(y)|a \rangle \langle b|,}
where $y$ is a longitudinal 24-dimensional coordinate and 
$T_{ab}$ is hermitian. 
As in \hklm, substituting into \rfgh\ reveals 
24-dimensional tachyons in the adjoint of $U(N)$. A similar
expansion gives $U(N)$ gauge fields. This is exactly the low-lying spectrum
of $N$ bosonic D23-branes. Higher mass open string states on the D25-brane
similarly descend to adjoint fields on the D23-branes, 
as in \refs{\hklm,\witnew}. 

\vskip.2in
\noindent{$U(\infty- N)$ Adjoint Fields}

Derivative terms in modes of the form $T_{jk}(y)|j \rangle \langle k|+h.c.$, 
where $j,k >N$ are projected out of the quadratic action 
because $C$ is proportional to 
$P_N$. Hence there are no propagating adjoint $U(\infty- N)$ fields.
In \refs{\hklm,\rms} the gauge field does not have a transverse profile 
(as is consistent with the boundary condition \dfgg )
and $C$ is proportional to the identity instead of $P_N$. 
In order to eliminate 
propagation of these modes, the additional assumption 
$f(T_{min})=0$ is required. Even then, 
if $f$ is quadratic or otherwise smooth about the minimum it may 
be set to one with a field redefinition. In these variables - which are the
natural ones for studying propagation -
propagating $U(\infty- N)$ tachyons reappear. 
In any case,  with the solution 
\soln\ the absence of such propagating modes is a natural 
consequence of the symmetries and no such additional assumptions 
about the $f$ prefactor or restrictions on field variables 
are necessary.

\vskip.2in
\noindent{$U(\infty- N)\times U(N)$ Bifundamental Fields}

We may also consider  
$U(\infty-N)\times U(N)$ bifundamental modes of the form 
$T_{ak}(y)|a \rangle \langle k|+h.c.$ where $a\leq N,
k>N$.\foot{In fact these modes can be gauged away or are eaten by the Higgs
mechanism \hklm, but then similar
comments pertain to the fluctuations of the gauge field. 
We consider the tachyon here
for notational simplicity.} 
Again, because of the projection operators in $C$, 
these modes do not acquire ordinary kinetic terms.   They do however have 
a nonvanishing quadratic action involving fixed matrices. 
Substituting into \rfgh\ we get 
\eqn\modes{S_{eff}(T_{ak})\sim 
{\rm
Tr}\Bigg[a^j a_j^{\dagger}T_{ak}^2 \Bigg]~~~~~~(j=2\cdots 13).}
\modes\ is the action for a charged particle in a magnetic field 
of strength ${1 \over \t}$. It has a discrete spectrum with 
spacing of order ${1 \over \t}$,  
rather than a spectra of continuous momenta. 
In particular, there are no bifundamental excitations 
with energies below $1 \over \t$.

Formally these modes disappear as the longitudinal 
noncommutativity $\t$ is taken to 
zero, however higher order corrections to the action \fti\ appear
to be suppressed by powers of $\apm \over \t$, and hence cannot be 
ignored at small $\t$. Hence we cannot make 
firm conclusions about the spectrum at small $\t$.

\newsec{Discussion}

In this paper we have proposed answers to two puzzles 
relating to the condensation of the open bosonic string tachyon
 (in the presence of a $B$ field)
\item{a.} Why are there no open string excitations at the bottom of 
the tachyon well for any value of $\T$ ?
\item{b.} Why is the condensed state at the bottom of the well  
independent of $\T$? 

We propose that 
as the tachyon rolls to its minimum, the gauge field also dynamically 
 rolls to its maximally symmetric value (with nonzero field strength),
 and the fully unbroken gauge invariance prohibits perturbative
propagation. 
This rolling is in part possible because, 
exactly at the bottom of the tachyon well, the coefficient of 
the Born-Infeld term in the action \fti\ vanishes, and there is no 
energy cost for changing a constant field strength.
Using the Seiberg-Witten change of variables, this maximally 
symmetric configuration with nonzero field strength and finite $\T$ 
can be reinterpreted as one with zero field strength and  
$\T=\infty$.

Restated, we propose that $\T$ (as set by the value of the commutative  
$\CF=F^\o+B$ at infinity) is effectively a dynamical variable 
that, regardless of its initial value, 
rolls to infinity in the process of tachyon condensation. 
The $\T$ independence of the tension and spectrum 
of our soliton is a consequence of this dynamical nature of $\T$.
If this proposal is indeed correct, it would be very interesting to 
understand in detail the dynamics that sends $\T$ to infinity, 
rather than any other (seemingly degenerate) value, as the tachyon
rolls to its minimum.

In closing we note that the $U(\infty)$ symmetry restoration
described here is obviously closely related to the symmetry 
restoration in the cubic formulation 
\cubic\ of Witten's open string field theory. It would be of interest to 
understand this connection in more detail.

\centerline{\bf Acknowledgements}
  We are grateful to J. Harvey, P. Kraus, F. Larsen,   
J. Maldacena, E. Martinec and A. Sen for useful discussions. 
Observations related to those of this 
paper have been independently 
made by the authors of reference \hklm\ (unpublished). 
This work was supported in part by DOE grant DE-FG02-91ER40654.

\listrefs
\end